\pdfoutput=1
\documentclass[sigconf]{acmart}

\usepackage{multirow}
\usepackage{longtable}
\usepackage{makecell,fancyvrb}
\usepackage{enumitem}

\newcommand{\uline}[1]{\underline{#1}}

\newcommand{\LANCE}{LANCE}

\newcommand{\AUCAD}{\textit{AUCAD}}
\newcommand{\AucadLog}{\textit{AucadLog}}

\newcommand{\link}[1]{~\href{#1}{#1}}
\newcommand{\artifact}{\link{https://github.com/aiopsplus/AucadLog}}

\AtBeginDocument{%
  }

\setcopyright{rightsretained}
\copyrightyear{2025}
\acmYear{2025}
\acmDOI{10.1145/3755881.3755889}

\acmConference[Internetware 2025]{the 16th International Conference on Internetware}{June 20--22, 2025}{Trondheim, Norway}
\acmBooktitle{the 16th International Conference on Internetware (Internetware 2025), June 20--22, 2025, Trondheim, Norway}
\acmISBN{979-8-4007-1926-4/25/06}

\begin{document}

\title{\AUCAD: Automated Construction of Alignment Dataset from Log-Related Issues for Enhancing LLM-based Log Generation}

\author{Hao Zhang}
\orcid{0009-0008-6205-3593}
\affiliation{%
  \institution{Software Institute, Nanjing University}
  \city{Nanjing}
  \state{Jiangsu}
  \country{China}
}
\email{hao-zhang@smail.nju.edu.cn}

\author{Dongjun Yu}
\orcid{0009-0009-3088-6213}
\affiliation{%
  \institution{Software Institute, Nanjing University}
  \city{Nanjing}
  \state{Jiangsu}
  \country{China}
}
\email{502023320014@smail.nju.edu.cn}

\author{Lei Zhang}
\orcid{0009-0001-3544-8636}
\affiliation{%
  \institution{Software Institute, Nanjing University}
  \city{Nanjing}
  \state{Jiangsu}
  \country{China}
}
\email{522023320200@smail.nju.edu.cn}

\author{Guoping Rong}
\orcid{0000-0003-4576-0524}
\authornote{Corresponding author.}
\affiliation{%
  \institution{Software Institute, Nanjing University}
  \city{Nanjing}
  \state{Jiangsu}
  \country{China}
}
\email{ronggp@nju.edu.cn}

\author{Yongda Yu}
\orcid{0000-0001-6713-2364}
\affiliation{%
  \institution{Software Institute, Nanjing University}
  \city{Nanjing}
  \state{Jiangsu}
  \country{China}
}
\email{yuyongda@smail.nju.edu.cn}

\author{Haifeng Shen}
\orcid{0000-0002-8221-981X}
\affiliation{%
  \institution{Faculty of Science and Engineering, Southern Cross University}
  \city{Bilinga}
  \state{Queensland}
  \country{Australia}
}
\email{haifeng.shen@scu.edu.au}

\author{He Zhang}
\orcid{0000-0002-9159-5331}
\affiliation{%
  \institution{Software Institute, Nanjing University}
  \city{Nanjing}
  \state{Jiangsu}
  \country{China}
}
\email{hezhang@nju.edu.cn}

\author{Dong Shao}
\orcid{0000-0001-6500-0341}
\affiliation{%
  \institution{Software Institute, Nanjing University}
  \city{Nanjing}
  \state{Jiangsu}
  \country{China}
}
\email{dongshao@nju.edu.cn}

\author{Hongyu Kuang}
\orcid{0009-0003-8702-2826}
\affiliation{%
  \institution{Software Institute, Nanjing University}
  \city{Nanjing}
  \state{Jiangsu}
  \country{China}
}
\email{khy@nju.edu.cn}

\renewcommand{\shortauthors}{Zhang et al.}


\begin{abstract}
  
Log statements have become an integral part of modern software systems. Prior research efforts have focused on supporting the decisions of placing log statements, such as where/what to log.
With the increasing adoption of Large Language Models (LLMs) for code-related tasks such as code completion or generation, automated approaches for generating log statements have gained much momentum. 
However, the performance of these approaches still has a long way to go. This paper explores enhancing the performance of LLM-based solutions for automated log statement generation by post-training LLMs with a purpose-built dataset. Thus the primary contribution is a novel approach called \AUCAD, which automatically constructs such a dataset with information extracting from log-related issues. Researchers have long noticed that a significant portion of the issues in the open-source community are related to log statements. However, distilling this portion of data requires manual efforts, which is labor-intensive and costly, rendering it impractical. Utilizing our approach, we automatically extract log-related issues from 1,537 entries of log data across 88 projects and identify 808 code snippets (i.e., methods) with retrievable source code both before and after modification of each issue (including log statements) to construct a dataset. Each entry in the dataset consists of a data pair representing high-quality and problematic log statements, respectively. With this dataset, we proceed to post-train multiple LLMs (primarily from the Llama series) for automated log statement generation. Both human and experimental evaluations indicate that these models significantly outperform existing LLM-based solutions, thereby validating the efficacy of our method for constructing a post-training dataset to enhance LLM-based log statement generation.

\end{abstract}

\ccsdesc[500]{Software and its engineering~Software creation and management}

\keywords{LLM, Log Statement Generation, Log-related Issues, Alignment}


\maketitle

\section{Introduction}
Log statements serve the essential role of documenting a series of discrete events within an active service, generating log data that is essential for software engineers to understand the operational status of software systems~\cite{Gatev2021}. Log data is indispensable for performing Operations and Maintenance (O\&M) tasks, including anomaly detection~\cite{Le2021, guo2024logformer} and root cause localization~\cite{Gu2023, Shi2023, Liu2024}. The quality of log data, which depends on the quality of log statements, has a significant impact on the effectiveness of these O\&M tasks~\cite{Le2022}. 
Consequently, improving the quality of log statements is vital for effective O\&M, and numerous studies have concentrated on enhancing log statement quality through best logging practices~\cite{Fu2014, pecchia2015industry}, particularly in the instrumentation of log statements~\cite{Chen2019, Zhang2022studying}.

With advancements in machine learning, various automation tools for logging practices have been proposed in both academia and industry. These tools aim to recommend specific elements in logging practices (e.g., log location~\cite{LogLocation}, log level~\cite{TELL}, and variables recorded in the logs~\cite{Liu2021}) or provide the capability to suggest complete log statements~\cite{JLLAR, Ding2022}. However, these tools are not yet sufficiently effective for the automated generation of log statements. Current mainstream logging practices remain largely manual, heavily relying on developers' experience and personal preferences~\cite{Kabinna2016, Gu2023TSE}. Consequently, many developers struggle to accurately log the desired information, leading to inconsistencies in the quality of the resulting log data~\cite{Rong2020ICSME} and subsequent O\&M tasks.

The emergence of Large Language Models (LLMs) has created promising prospects for research and applications in related fields \cite{LinICSE2024Brief}. 
Researchers are investigating the advanced coding capabilities of LLMs to directly recommend complete log statements through a prompt engineering approach, such as UniLog~\cite{UniLog} and SCLogger~\cite{LiFSE24GoStatic}. 
\LANCE~\cite{LANCE} and FastLog~\cite{xie2023generate} represent another technical approach -- fine-tuning -- commonly employed in research concerning LLM-enabled software practices~\cite{Fan2023LLMSESurvey}. This approach is particularly prevalent in studies leveraging open-source LLMs, such as the Llama series~\cite{llama31modelcard}. 
Nevertheless, post-training of LLMs, which includes both fine-tuning and alignment stages, for automated log statement generation remains an underexplored area, particularly with respect to the alignment phase. A critical challenge in this domain is the scarcity of high-quality datasets for post-training. In the case of alignment, such a dataset necessitates the inclusion of both positive and negative samples. While it is well-established that a significant portion of issues in the open-source community pertains to log statements, constructing a dataset from these issues demands extensive manual efforts, rendering it impractical.

This paper addresses this gap by proposing \textbf{\AUCAD} (\textbf{AU}tomated \textbf{C}onstruction of \textbf{A}lignment \textbf{D}ataset), a novel method that achieves automated construction of high-quality alignment datasets specifically designed for enhancing LLM-based log statement recommendation through post-training. Our key innovation lies in establishing explicit correspondence between log statement code contexts and their evolutionary changes through systematic analysis of log-related issues in open-source communities. 
A pivotal component of this method is the construction of a distilled, structured dataset composed of high-quality log statements well-recognized within the open-source community. The objective of this dataset is to improve the performance of post-training LLMs in generating log statements. The proposed method collects log-related issues from the Apache open-source community, from which log statements and their corresponding code context -- post-resolution in open-source software systems -- are automatically extracted and structured to create a distilled alignment dataset known as \AucadLog.
This dataset is considered to be high-quality for two main reasons. First, the contextual code snippets related to the log statements have garnered significant attention from the community. Second, the final form of the log statements and contextual code snippets reflects the approval of senior developers, indicating the prominence and quality of these log statements in comparison to others. This construction enables LLMs to learn and master better logging practices from the open-source community, thereby facilitating the recommendation of high-quality log statements.

The evaluation of the dataset indicates that the dataset constructed using the \AUCAD\ method ensures the relevance and effectiveness of positive examples, thereby guaranteeing both the validity and high quality of the dataset. This finding also underscores the scalability of the \AUCAD\ method. Furthermore, implementation results across various alignment algorithms, including \textit{DPO}~\cite{DPO}, \textit{cDPO}~\cite{cDPO}, and \textit{SimPO}~\cite{SimPO}, demonstrate that the \AucadLog\ dataset constructed through the \AUCAD\ method effectively supports the alignment of LLMs. They also exhibit significant performance improvements in enhancing the capability of open-source LLMs to automatically generate log statement.

The main contributions of this paper are as follows:
(1) We propose \textbf{\AUCAD\ Method}, a novel method for the automated construction of an alignment dataset with both positive and negative log statements. 
By automatically extracting and structuring log statements with contextual code, \AUCAD\ ensures high-quality datasets that enhance log generation performance and scalability.
(2) We contribute \textbf{\AucadLog\ Dataset}, a custom-built dataset through the \AUCAD\ method that comprises both high-quality and problematic log statements as positive and negative samples for the research community to explore the alignment of LLMs.

\section{Related Work}\label{section:related}

This section discusses related works on the recommendations of log statements and the adoption of LLMs for automated software engineering tasks.

\subsection{The Generation/Recommendation of Log Statements}

Closely related to code generation, there has recently been a surge in research of log statement instrumentation~\cite{chen2021survey}. 
In recent years, researchers have conducted extensive empirical studies on log statements in open-source software, constructing diverse log statement datasets for systematic analysis. For instance, Hassani et al.~\cite{hassani2018studying} performed a manual investigation of hundreds of issues, identifying seven root causes of log-related problems. Chen et al.~\cite{LCC-SZZ} proposed the LCC-SZZ to automatically detect Logging-Code-Issue-Introducing Changes (LCII) from software version histories, tracing the evolutionary patterns of log statements to pinpoint issue origins. While these empirical studies provide critical insights for enhancing logging practices and understanding log maintenance patterns, they predominantly focus on recording metadata of code changes (i.e., log statement modifications) without fully leveraging the rich contextual information pervasive in open-source communities. Moreover, these datasets cannot be directly applied to log statement recommendation tasks. 

The advent of numerous AI applications for code generation tasks has spurred the gradual integration of AI techniques into log-related practices~\cite{li2021deeplv, LANCE, mastropaolo2024log, ALS, ELogger}. For a considerable period, studies in this area were confined to addressing specific sub-problems of log statement instrumentation, e.g., the position (where to log) and the content (what to log) of log statements~\cite{Gu2023TSE}. For instance, Zhang et~al.~\cite{LogLocation} proposed DeepLog, a model that recommends log statement positions using a double-branched neural network model pre-trained with semantic and syntactic features extracted from abstract syntax trees. Zhu et~al.~\cite{LogAdvisor} introduced LogAdvisor, which employs feature selection, classifier learning, and noise handling techniques to assist developers in determining where to place log statements.

To determine the content of log statements, 
Liu et~al.~\cite{Liu2021} introduced a method where each program token is labeled as a vector of real numbers through RNN and self-attention layers. This approach predicts the probability of each token being logged and recommends the variables that are suitable to be logged. 
Ding et~al.~\cite{Ding2022} proposed LoGenText, a model that recommends log texts using neural machine translation models. 
Gholamian et~al.~\cite{gholamian2021log} employed log-aware clone detection to achieve automated log position and description prediction. While the verbose level determines the information captured by log statements, Liu et~al.~\cite{TELL} put forth TeLL, a model that leverages graph neural networks to encode features, thereby facilitating the recommendation of log levels. Li et~al.~\cite{li2021deeplv} proposed DeepLV, which utilizes recurrent neural networks to predict appropriate verbose levels.

\subsection{Adoption of LLMs for Automated Software Engineering Tasks}

With the rapidly increasing ability of content comprehension and generation, the emergence of LLMs is regarded as a major disruption to software development~\cite{huang2024generative}. Much of the research surrounding LLMs is focused on code-related tasks, such as code generation, completion, summarization, review, and testing~\cite{zhang2023unifying}.
Take coding task as an example. There are many solutions adopting LLMs in coding, for example, Cursor~\footnote{\link{https://www.cursor.com/}}, 
GitHub Copilot~\footnote{\link{https://github.com/features/copilot}}, 
and Alibaba's TONGYI Lingma~\footnote{\link{https://tongyi.aliyun.com/lingma/}}
have greatly improved developers' efficiency~\cite{xing2023prompt}. Meanwhile, there are also many examples of applying LLMs for noncoding purposes such as code review (e.g., LLAMA-Reviewer~\cite{lu2023llama}), code summarization~\cite{sun2023automatic, ahmed2023improving, geng2024large}, and software testing~\cite{wang2024software}.

Despite the extensive applications of LLMs across various domains in software engineering, their utilization in log statement generation is surprisingly sparse. To the best of our knowledge, there are only a handful of studies on this specific topic, e.g.,  \LANCE\ by Mastropaolo et~al.~\cite{LANCE}, FastLog by Xie et~al.~\cite{xie2023generate}, LEONID by Mastropaolo et~al.~\cite{mastropaolo2024log}, and UniLog by Xu et~al.~\cite{UniLog}, and SCLogger by Li et~al.~\cite{LiFSE24GoStatic}, etc.
\LANCE~\cite{LANCE} is a model that attempts to generate entire log statements in an end-to-end manner using the T5-small model. 
FastLog~\cite{xie2023generate} is an end-to-end approach for log statement generation and insertion based on the fine-tuned PLBART~\cite{PLBART} model.
LEONID~\cite{mastropaolo2024log} introduces minor modifications to the \LANCE\ model, including the use of a larger pre-training dataset and support for multiple log statements, leading to marginal improvements in log statement recommendation. 
The research proposed by Xu et~al.~\cite{UniLog} employs the Codex as the core of UniLog, which has a greater number of parameters. This model leverages the in-context learning capability of the LLM for warm-up training and achieves superior log statement recommendation results on the \LANCE\ dataset. However, it is evident that the limited amount of research conducted so far has not adequately addressed the practical challenges of automatically generating high-quality log statements within source code. 
SCLogger\cite{LiFSE24GoStatic} utilizes the principles of prompt engineering to organize the context in alignment with a Chain-of-Thought (CoT) strategy, using refinement prompts to integrate comprehensive variable type information into the preliminary log statement.
In fact, a recent empirical evaluation\cite{Li2024LogBench} suggests that, possibly due to data leakage, the performance of most of the existing LLMs in log statement generation may be overestimated, further revealing the necessity for an in-depth research on this topic.

In summary, LLMs have been adopted in nearly all processes in software engineering~\cite{Fan2023LLMSESurvey}. As these processes often involve specialized domain knowledge and there are concerns about privacy, security, and other factors~\cite{kostova2020privacy}, fine-tuning LLMs presents a prevailing approach to applying LLMs to software engineering tasks~\cite{lin2024data}. The prominence of using LLMs for code generation-related tasks suggests their potential for the task of automated log statement generation.

\begin{figure*}[htb]
    \centering
    \includegraphics[width=0.9\textwidth]{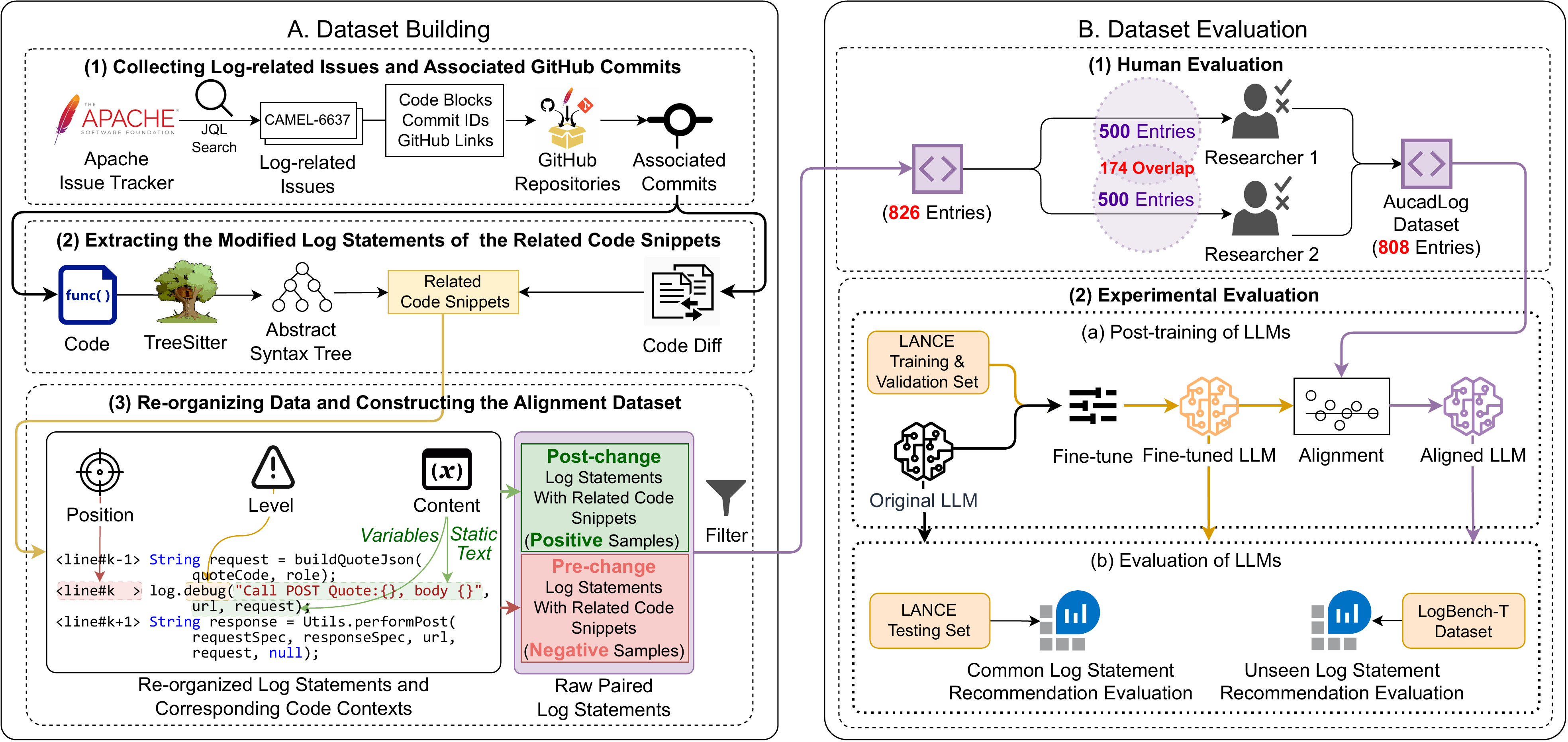}
    \caption{The \AUCAD\ method for automated construction of an alignment dataset with log statements.}
    \label{fig:overall}
\end{figure*}

\section{Methodology Overview}\label{section:method}
In this section, we outline \AUCAD, which constructs a distilled, structured dataset with log statements from log-related issues in the open-source community for the alignment of open-source LLMs.

Generally, the recommendation of log statements needs the following three sub-tasks: 
(1) \textbf{Position}: recommend the optimal position of log statements within the source code; 
(2) \textbf{Level}: recommend the appropriate verbosity level for the log statement (e.g., TRACE, DEBUG, INFO, WARN, or ERROR); 
(3) \textbf{Content}: generate the content/variables of the log statement to capture the required information.
Therefore, we collected log statements from the resolved log-related issues in the open-source community and organized the dataset to specifically address these three sub-tasks. The dataset provides structured information to support the recommendation of optimal positions, appropriate verbosity levels, and meaningful content of log statements, effectively guiding the automated generation of comprehensive log statements. 
Quality of log statements is a distinctive feature of our work that distinguishes itself from similar works~\cite{LogAdvisor,LANCE}, which typically collect log statements present in the source code of the open-source software without considering their evolution over the software development lifecycle. We paid particular attention to the log statements before and after the modifications in the log-related issues, which were used as negative and positive samples of logging practice, to construct a distilled dataset for the post-training of open-source LLMs.

Fig.~\ref{fig:overall} illustrates the \AUCAD\ method, which consists of two primary stages: 
(A) \textbf{dataset building} (Section~\ref{section:method-dataset}), 
and (B) \textbf{dataset evaluation} (Section~\ref{section:method-evaluation}). Each stage is described in detail in the subsequent sections.

In the first stage, we design a three-step process to collect and filter log-related issues from the Apache Issue Tracker. 
We extract code snippets from associated GitHub commits to build a distilled, structured dataset, called \AucadLog, containing log statements and their associated context code after modifications in these issues, so as to build paired samples with both positive and negative examples of log statements. 

In the second stage, we evaluate the dataset through (1) \textbf{human evaluation} (Section~\ref{section:method-evaluation-human}) and then (2) \textbf{experimental evaluation} (Section~\ref{section:method-evaluation-experimental}). For \textbf{human evaluation}, two independent researchers assess code pairs for relevance to the corresponding issue's intent, identifying high-quality code diffs verified by human judgment. The \textbf{experimental evaluation} defines the optimization task, followed by fine-tuning Llama 3.1 and Magicoder (leading open-source and code-specific LLMs) to develop exemplar models with enhanced log-statement generation capabilities. Subsequently, we apply preference alignment algorithms with the \AucadLog\ alignment dataset to further refine the LLMs by alignment. Finally, we evaluate the original LLM, fine-tuned LLM, and aligned LLM in their ability to generate log statements for both common and unseen code contexts.

\section{Dataset Building} \label{section:method-dataset}

There are a large amount of log-related issues in the open-source community, offering valuable insights for logging practices. Compared to log statements and their contexts obtained directly from the source code of the open-source software, those reviewed and discussed in the resolved issues tend to be of higher quality. Some existing studies~\cite{Claes202020-MAD} have paid attention to log-related issues and realized the significance and value embedded in this part of the data. The presence of a log-related issue often indicates suboptimal practices in the original logging approach, whereas the log statements revised through discussion are more reliable to be a good reference. Moreover, many attempts at fine-tuning LLMs have shown a significant positive effect of high-quality fine-tuning data on model performance~\cite{chang2024data, tajwar2024preference, yang2024fine}.
Therefore, we chose to construct a high-quality distilled dataset by extracting the log statements after the modifications in the log-related issues, and the associated contextual code snippets.

As depicted in Fig.~\ref{fig:overall}, we developed \AUCAD\ for constructing a high-quality log statement dataset \AucadLog\ from the Apache Software Foundation and GitHub. This method comprises three steps: 
(1) collecting a large number of log-related issues from open-source software projects of the Apache Software Foundation and retrieving their associated GitHub commit information, 
(2) extracting the modified log statements and the relevant code snippets as our distilled data,
(3) structuring the distilled data to construct the fine-tuning dataset.
Detailed steps are elaborated below.

\subsection{Step 1: Collecting Log-related Issues and Associated GitHub Commits}

Obviously, the more active the open-source project community means that there are more opportunities to find suitable log-related issues and a relative wealth of discussion information. With this in mind, we refer to the annual reports of the Apache Software Foundation~\footnote{\link{https://www.apache.org/foundation/docs/FY2023AnnualReport.pdf}} to establish the project candidate pool, and finally select 88 project teams based on the number of issues ($\ge 100$) and participants (\textit{watch} or \textit{fork} or \textit{star}$\ge 500$).

To efficiently gather issue data from the Apache Software Foundation, we utilized the Apache Issue Tracker~\footnote{\link{https://issues.apache.org/}}, which manages issues, tasks, and enhancement requests for Apache projects. This system, which includes both Bugzilla 
and Jira, 
enables community members to report and discuss various issues. We favored Jira for its advanced Jira Query Language (JQL), allowing precise issue filtering. 
We streamlined the data retrieval process by focusing on issue summaries, which concisely encapsulate the content, and employed keywords related to logging to filter relevant issues, e.g., \emph{log}, \emph{logger}, \emph{print} and \emph{logging}, etc. However, to avoid irrelevant results, we excluded phrases not associated with logging practice, typically including \emph{log in}, \emph{log out}, \emph{blue print} or \emph{print command}, etc. 
We targeted four issue types -- \emph{Bug, Dependency, Dependency upgrade,} and \emph{Improvement} -- to extract meaningful log patterns. Moreover, as only resolved issues were considered to ensure data quality, we used the \texttt{resolved date} field as a filter. The time range for our data extraction was from January 1, 2002, to December 31, 2024, covering almost all the log-related issues from the Apache projects. 

After collecting the issues, we further processed and analyzed the raw data to establish associations with GitHub code repositories. Specifically, we searched the issue descriptions and comments for links to GitHub or GitBox, which include commits, code snippet references, GitHub issues, and pull requests. As all Apache repositories on GitBox are available on GitHub, we utilized the GitHub REST API~\footnote{\link{https://docs.github.com/en/rest}} to develop a crawler that retrieved the corresponding commit's code change information (i.e., code diff) and the modified code files. This allowed us to restore the committed code context using the diff and the changed code. For code snippet references, we obtained the complete content of the corresponding file after the commit. For related GitHub issues and pull requests, we identified the last valid commit in the pull request timeline. Through this process, we established the relationship between log-related issues and their associated GitHub commit information.

\subsection{Step 2: Extracting the Modified Log Statements and the Relevant Code Snippets}

Next, we extracted the modified log statements and related code snippets from the gathered information to serve as the distilled data for our dataset. We utilized tree-sitter~\footnote{\link{https://tree-sitter.github.io/tree-sitter/}} to perform static analysis on the code files obtained from GitHub and constructed Abstract Syntax Trees (ASTs) to extract the methods within the code files. Additionally, we used a combination of ASTs and regular expressions to extract all log statements present within the methods. 
We then analyzed the code diff to generate changed log statements before and after the modification. We automated the analysis of code diffs to extract all log statement changes associated with each issue, identifying log statements flagged by issues and subsequently modified along with their related context. Post-change log statements in the discussions on issues are treated as examples of better logging practices, i.e., positive examples. Conversely, the original, pre-change log statements -- identified as problematic -- are considered poor logging practices, serving as negative examples. It is noted that to ensure the quality of the constructed dataset, we extracted the complete method code containing the target log statement as the code context, including only the Java methods. In particular, the entire conditional structure was extracted if a conditional structure contained only the log statement.

\subsection{Step 3: Structuring Data and Constructing the Dataset}

Finally, we structured these log statements and their corresponding code contexts to build a high-quality distilled, structured dataset \AucadLog\ for the post-training of LLMs.

The dataset contains the most recent versions (i.e., the committed versions) of log statements, reflecting better logging practices accepted by the open-source community following issue discussions and revisions.  Additionally, it includes the original, pre-change versions of these log statements, serving as examples of poor logging practices for comparison and alignment. 

Additionally, to ensure the dataset quality, we applied automated filtering approaches based on rules and code context AST analysis to \AucadLog\ Dataset for the further post-training LLMs. 
This process removed 
cases with trivial inconsistencies between expected positive and negative examples (e.g., where positive and negative examples differed only in code indentation), and mitigated potential data leakage in the testing set.
In addition, considering that logging changes may occur concurrently with modifications to the functional code, we filtered out items that have code contexts closely related to the functional code.

\section{Dataset Evaluation} \label{section:method-evaluation}

To validate the effectiveness of the \AucadLog\ dataset constructed via \AUCAD, we conduct multi-dimensional evaluations from both human and experimental perspectives.

\subsection{Research Questions}
To evaluate the quality and comprehensibility of the dataset constructed using \AUCAD, along with its potential implications for log statement recommendation, we propose the following two research questions:

\begin{itemize}
    \item \textbf{RQ1}: To what extent can the \AUCAD\ method correctly identify positive and negative log statements?
    \item \textbf{RQ2}: How can the \AucadLog\ enhance the capabilities of LLMs in log statement recommendation through post-training?
\end{itemize}

\textbf{RQ1} seeks to comprehensively validate the dataset developed through the \AUCAD\ method, ensuring it correctly captures both good log statements (positve examples) and bad log statments (negative examples) as expected. 
Following this, \textbf{RQ2} assesses the practical utility of this dataset in the post-training of LLMs, utilizing various base models and alignment algorithms.

\subsection{Human Evaluation} \label{section:method-evaluation-human}

Based on the automatically filtered \AucadLog\ dataset, two independent researchers operated separately to carefully identify if the related code diff pairs are relevant to their corresponding issues. 

For the randomly shuffled 826 entries, each researcher was assigned 500 of them to ensure that every diff was examined, with 174 entries selected to be labeled twice for measuring inter-rater reliability by computing Cohen's kappa coefficient\cite{cohen1960coefficient} to check if the two researchers mark the diffs consistently.

\subsubsection{Inclusion Criteria.}
The criteria for determining the relevance of \AucadLog\ entries to their associated issues are as follows:
\begin{enumerate}
\item \textbf{Relevant entries}: Diffs are deemed as relevant entries if they align clearly with the issue's stated objectives, such as directives for log level adjustments, addition or removal of logged variables, adaptation to a new logging framework, or the adoption of parameter-based logging styles. Obviously, this relevance is a semantic level relevance that requires the evaluator to read and understand both the diff and the corresponding discussion around the issue.
\item \textbf{Non-relevant entries}: Others.
\end{enumerate}

\subsubsection{Settings.}
To manually validate the quality of the \AucadLog\ dataset, two separate researchers are participated. Each researcher has development experience in the industries whose main development language is Java. During the evaluation, the following information is provided:
\begin{enumerate}

\item\textbf{Related Issue Title}: the title of the related issue, in the format like \textit{[FLINK-1220] Make INFO logging more verbose}.
\item\textbf{Related Issue URL}: URL of the issue in Apache Issue Tracker.
\item\textbf{Related Issue Description}: original issue description. 
\item\textbf{Related Issue Comments}: comments of the related issue, including the commentor's name and the content of the comment.
\item\textbf{Origin Code}: code before modification of the referenced log issue.
\item\textbf{Accepted fixes}: code after modification of referenced log issue.
\end{enumerate}

The evaluators are supposed to read the source code both before (i.e., \textbf{Origin Code}) and after (i.e., \textbf{Accepted fixes}) modification as well as the \textbf{Related Issue Comments} to determine relevance using the \textit{Inclusion Criteria}. They are encouraged to use other information listed above to confirm the judgment.
To guarantee the consistency, the two researchers worked independently on their assigned 500 code diffs, among which 174 diffs were shared across the two researchers, which is used to 
calculate the Cohen's kappa coefficient.

\subsection{Experimental Evaluation} \label{section:method-evaluation-experimental}

In the post-training phase, which includes fine-tuning or alignment, these models are further refined to enhance their capacity to apply the acquired knowledge~\cite{zhao2023survey}. 
In this paper, we utilize the dataset constructed from discussions and modifications of log-related issues during the post-training phase of LLMs, particularly in alignment.

\subsubsection{Selection of base LLMs \& baselines.}
Given that most commercial LLMs are either closed-source or provide access exclusively through API calls, concerns over data security (particularly in enterprises) and cost considerations made open-source LLMs the preferred choice for fine-tuning and alignment. Specifically, we utilized Llama~3.1~\cite{llama31modelcard} and Magicoder (based on \textsc{CodeLlama})~\cite{wei2023magicoder} as our base models. These models have undergone extensive pre-training on both natural language and code datasets. The Llama series is one of the most prominent LLMs in the open-source community. Meanwhile, 
Magicoder (based on \textsc{CodeLlama})~\cite{wei2023magicoder}, as a specialized open-source LLM for code-related tasks. Magicoder has been extensively pre-trained on code datasets, making it particularly effective for code understanding and generation. 
Furthermore, to accommodate GPU resource, service availability as well as the specific requirements of recommendation tasks, we selected instruction-tuned versions of these models with relatively smaller parameter sizes. Specifically, we used \textit{Meta-Llama-3.1-8B-Instruct} and \textit{Magicoder-S-CL-7B} as the base models for post-training.

Since we intend to compare the alignment effects of our dataset at a similar scale, we set the following criteria to select benchmark apporoaches.
(1) The selected approaches must rely on LLM technologies;
(2) The selected approaches must be able to use the dataset in our study.

As the result, we identified the following approaches:
\begin{enumerate}
    \item \LANCE~\cite{LANCE} method  utilizes a pre-trained and fine-tuned T5 model~\cite{T5} for recommending log statements in Java code.  
    \item FastLog~\cite{xie2023generate} is an end-to-end approach for log statement generation and insertion based on the fine-tuned PLBART model~\cite{PLBART}.       
    \item UniLog~\cite{UniLog}, which is based on prompt engineering using the LLM of the same parameter scale as the backbone.
\end{enumerate}
It is noteworthy that although one approach SCLogger~\cite{LiFSE24GoStatic} also represents a recent prompt engineering-based approach, the data different structure applied by SCLogger (methods in project) and our approach (independent methods) make the comparison very limited sense.

\begin{table}[htb]
\caption{Datasets used by experiments.}
\label{tab:datasets}
\footnotesize
\begin{tabular}{|l|l|r|l|l|}
\hline
\multicolumn{1}{|c|}{\textbf{Dataset}} & \multicolumn{1}{c|}{\textbf{Purpose}} & \multicolumn{1}{c|}{\textbf{Size}} & \multicolumn{1}{c|}{\textbf{Data Composition}}  \\ \hline
\textbf{LANCE} & \begin{tabular}[c]{@{}l@{}}Fine-tuning; \\ Evaluation\end{tabular} & 131,662 & Java methods \\ \hline
\textbf{\AucadLog} & Alignment & \begin{tabular}[c]{@{}l@{}}808\end{tabular} & \begin{tabular}[c]{@{}l@{}}Positive-negative example pairs \\ from log-related issues\end{tabular} \\ \hline
\textbf{LogBench-T} & Evaluation & \begin{tabular}[c]{@{}l@{}}6,849\end{tabular} & \begin{tabular}[c]{@{}l@{}}Transformed unseen code\end{tabular}  \\ \hline
\end{tabular}
\end{table}

\subsubsection{Datasets.}
The experiments mainly use three datasets (as shown in Table~\ref{tab:datasets}): \LANCE\ dataset~\cite{LANCE}, which supports initial fine-tuning of the model, our \AucadLog\ dataset, which provides unique pairs of positive examples of improved log statements and negative examples of poor logging practices, and LogBench-T dataset~\cite{Li2024LogBench}, which includes unseen code for most LLMs to evaluate the model.
While existing empirical studies have constructed log-related datasets \cite{hassani2018studying,LCC-SZZ}, most merely catalog predefined static features of log statements (e.g., verbosity levels, message changes) without preserving their contextual code dependencies. This critical limitation renders such datasets incompatible with log statement generation tasks, thereby justifying their exclusion from our experimental framework.
In the post-training phase, the goal is to enhance the capability of the LLMs to accomplish the log statement generation task. Note that since our dataset consists of only 808 entries, which, according to common practices (considering that \LANCE\ dataset that used by \cite{xie2023generate} contains 131,662 entries),
is insufficient to support effective fine-tuning on its own. Therefore, we employed the \LANCE\ dataset for initial fine-tuning, followed by alignment using our \AucadLog\ dataset. Presenting both the fine-tuning results and the outcomes after alignment allows for a comparative analysis that aids in understanding the significance of our data.

Notably, our \AucadLog\ dataset \textbf{uniquely} offers both positive examples of improved log statements (after the changes) and negative examples of poor logging practices (before the changes), thereby creating natural pairs of data -- \textit{winning and losing responses} in \textit{the same context}. In contrast to conventional log-related datasets~\cite{hassani2018studying,LANCE,LCC-SZZ}, \AucadLog\ establishes contextual associations between code structures and contrasting log statement pairs. This distinct characteristic allows us to effectively perform alignment. In this work, we leverage such data to further enhance the fine-tuned model using alignment techniques in order to train the LLM to generate higher-quality log statements.

In the evaluation, we selected the testing set from the \LANCE\ dataset and the LogBench-T dataset~\cite{Li2024LogBench} to assess the model's performance. In particular, the LogBench-T dataset proposed by Li et al.~\cite{Li2024LogBench} includes unseen code for most LLMs, addressing potential data leakage problems during the pre-training phase of open-source LLMs that could affect the final evaluation results.

\subsubsection{Post-training.}
\textit{\uline{Fine-tuning}.}
Fine-tuning an LLM can significantly enhance its capabilities under zero-shot conditions~\cite{WeiBZGYLDDL22Finetuned}, forming the foundation for equipping the model with domain-specific expertise. During the fine-tuning phase, they are refined and guided by a smaller, task-specific dataset to enhance their performance on downstream tasks~\cite{Yang2024Harnessing}.

However, an LLM typically has a large number of parameters, and as such, it is resource-intensive to fully fine-tune it. To save computational and storage costs, we adopt low-parameter fine-tuning (i.e., LoRA~\cite{Edward2022LoRA}). 
We adopt a rank of $r=8$ as the parameter for fine-tuning, which enables us to complete the fine-tuning procedure with merely 0.05\% of the model's parameter ratio.
To further reduce the training cost and improve the convergence speed, we adopt the more efficient optimizer Lion (Evo\textbf{L}ved S\textbf{i}gn M\textbf{o}me\textbf{n}tum)~\cite{NEURIPS2023_Lion}, thereby achieving model fitting in fewer training iterations while also reducing GPU memory requirements.

\textit{\uline{Alignment}.}
Unlike fine-tuning, alignment includes both positive examples of desirable responses and negative examples of undesirable responses. This dual approach helps prevent the model from generating harmful or inappropriate outputs.

An effective approach to aligning LLMs is Reinforcement Learning from Human Feedback (RLHF)~\cite{RLHF}. While classical RLHF has demonstrated strong performance, its multi-stage process introduces several optimization challenges, such as the need to train a reward model followed by the optimization of a policy model to maximize that reward. Recently, researchers have investigated simpler offline algorithms~\cite{DPO, SimPO, cDPO, IPO, TDPO, zhao2023slic}. In our study, we selected three alignment algorithms -- \textit{DPO} (Direct Preference Optimization)~\cite{DPO}, \textit{cDPO} (conservative \textit{DPO})~\cite{cDPO}, and \textit{SimPO} (Simple Preference Optimization)~\cite{SimPO} -- to align the model, leveraging the advantages of paired data derived from code changes linked to issues in open-source community logs. Unless mentioned in the paper, all hyper-parameters adopt the default settings in the original papers\cite{DPO, cDPO, SimPO}.

\textit{\uline{Prompting strategy}.}
In this work, the fine-tuning experiments use the prompting strategy of a regular \textit{code completion scheme}, as outlined below. In this approach, the task of generating log statements is framed as a code completion task. This is achieved by constructing zero-shot prompts that instruct the model with complete code snippets.

\begin{center}
\fbox{\parbox{0.85\columnwidth}{
\small
\textit{Recommend the optimal log statements in the following given codes.
You need to output the full code with optimal log statement inserted, and do not explain the reason.}

\textit{Code:}

\texttt{\`{}\`{}\`{}\{code\}\`{}\`{}\`{}}
}}
\end{center}

Specifically, for \LANCE\ and FastLog, which do not support the prompt engineering approach, we utilized the model-specific input format for evaluation (i.e., only the \texttt{code} part above as the model input without the rest accompanying instructional text).

\subsubsection{Evaluation metrics.}
In our evaluation, we selected the typical metrics adopted by the previous studies~\cite{li2021deeplv, LANCE, UniLog} to evaluate the position, verbosity level, and message in generated log statements. Details of the metrics are elaborated below.

\textit{Position Accuracy (PA)} is calculated as follows. If the position of the generated log statement matches the position of the actual log statement of ground truth in the source code for testing, \textit{PA} = 1 (indicating a successful recommendation); otherwise, \textit{PA} = 0 (indicating an unsuccessful recommendation). 

\begin{figure}
    \centering
    \includegraphics[width=0.85\linewidth]{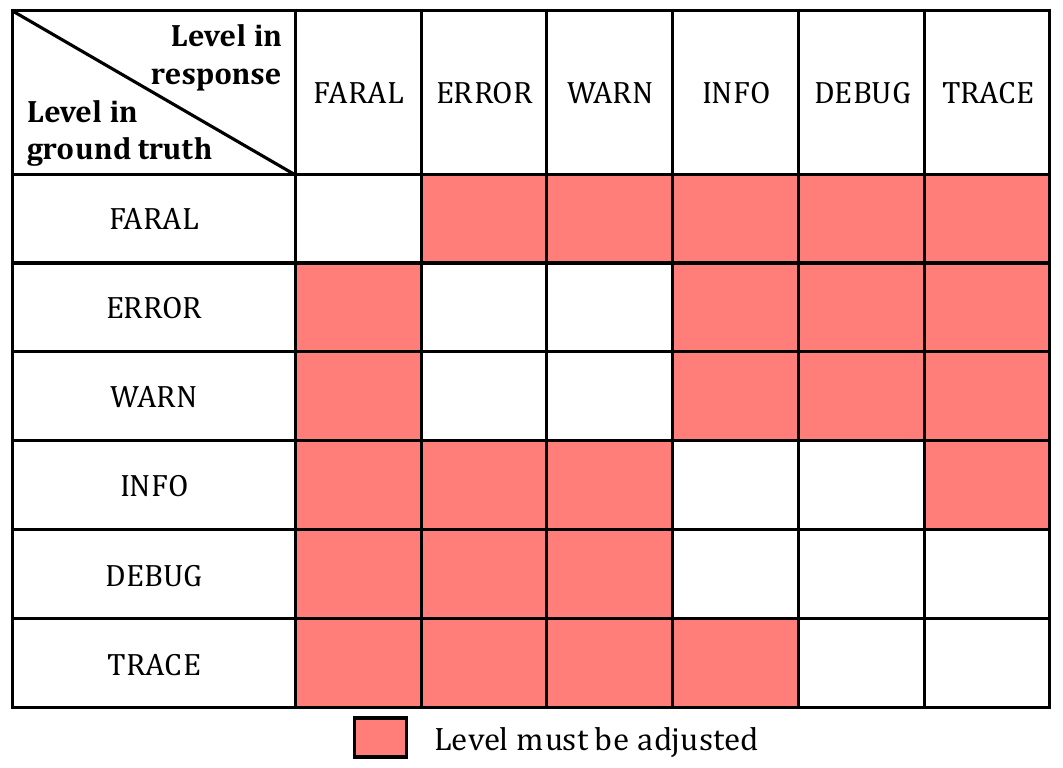}
    \caption{Level must be adjusted between the ground truth and responses when calculating Adjusted Level Accuracy.}
    \label{fig:adj-la}
\end{figure}

\textit{Level Accuracy (LA)} is calculated as the percentage of correctly predicted log levels, which involves comparing the predicted log levels against the actual log levels and determining how many predictions are correct.
However, considering that there is no strict standard for log levels, and that different logging practitioners might have varying habits for assigning levels (e.g., the INFO and DEBUG level logs may not make a significant difference in real logging practice), it seems unfair to directly compare the log levels recommended by the model with the ground 
truth. Therefore, we propose \textit{Adjusted Level Accuracy} (\textit{Adj. LA}), which only examines specific log levels that must be adjusted when the model-recommended log level does not match the ground truth. 
The rationale behind \textit{Adj. LA} is very straightforward, Fig.~\ref{fig:adj-la} shows all the levels that need to be adjusted, which constructed from large-scale log-related issues~\cite{hassani2018studying}. Firstly, the log verbosity levels are sequentially encoded from the lowest level TRACE (0) to the highest level FATAL (5). If recommended log statement falls into any of the levels in the corresponding right-hand column for a certain verbosity level (the left column), we consider it to be a wrong recommendation (i.e., the verbosity level must be adjusted), otherwise, we consider it to be a correct recommendation. For example, if the ground truth is the INFO level and the recommendation is DEBUG, we still consider it to be a correct Level recommendation.
Then \textit{Adj. LA} is calculated as the percentage of correct level recommendation to all the recommended results in this sense.


\textit{Message Accuracy (MA)} calculates the proportion of words in the response that are also present in the real text in ground truth as $\text{\textit{MA}} = {|W_p \cap W_t|}/{|W_p|}$, where $W_p$ and $W_t$ represent the constituent words split in model response and the real text in ground truth, respectively.

\textit{Text Accuracy} is calculated using \textit{BLEU}~\cite{BLEU} in NLP to evaluate the performance of the message generated by different models. \textit{BLEU} method evaluates the similarity of the text by comparing the candidate text with one or more reference texts, which is broadly used in the machine translation task. In particular, we adopt the \textit{BLEU-DM} variant~\cite{shi2022evaluation}, the sentence-level \textit{BLEU} with no smoothing method included, to obtain a more precise evaluation of the log messages in a more concise way, maintaining the integrity and fidelity of the information in the log statements. To be specific, the \textit{BLEU-DM} score is calculated based on n-gram precision and brevity penalty, 
i.e., $\text{\textit{BLEU-DM}} = \text{BP} \cdot \exp\left(\sum w_n \log p_n\right)$,
where $p_n$ is the n-gram precision (i.e., the proportion of n-grams in which the candidate text matches the reference text), $w_n$ is the n-gram weight, 
and $\text{BP} = \max\left(\exp{(1 - r / c)}, 1\right)$ 
is the brevity penalty for penalizing too-short candidate texts,
where $c$ is the length of the candidate text, and $r$ is the length of the reference text.

\textit{Variable Accuracy} evaluate the   
accuracy of variables recorded in the log statements, we denote the variables in each generated log statement as $V_p$ and the variables in each actual log statement of ground truth as $V_t$. Therefore, we calculated the \textit{Variable Precision} \(VP = {|V_p \cap V_t|}/{|V_p|}\), \textit{Variable Recall} \(VR = {|V_p \cap V_t|}/{|V_t|}\), and \textit{Variable F1 score} \(VF1 = 2 \times {VP\times VR}/(VP + VR)\) as metrics.
Note that if the same variable is used differently in log statement, they will be treated as different variables.

\begin{table}[htb]
\caption{The experiment settings.}
\label{tab:experiment-settings}
\footnotesize
\centering
\begin{tabular}{|l|l|l|}
\hline
\multicolumn{1}{|c|}{\textbf{Method}} & \multicolumn{1}{c|}{\textbf{Training \& Validation Set}} & \multicolumn{1}{c|}{\textbf{Testing Set}} \\ \hline
Original Model & N/A & \multirow{7}{*}{\begin{tabular}[c]{@{}l@{}}\LANCE\ testing set,\\ LogBench-T\end{tabular}} \\ \cline{1-2}
UniLog-w/ warmup & \begin{tabular}[c]{@{}l@{}}\LANCE\ training set (as examples) \\ and validation set (for warmup)\end{tabular} & \\ \cline{1-2}
FT & \multirow{4}{*}{\begin{tabular}[c]{@{}l@{}}
\textbf{Fine-tuning} (FT for short):  \\
\LANCE\ training \& validation set \\
\textbf{Alignment}: \AucadLog
\end{tabular}} &  \\ \cline{1-1}
FT+\textit{DPO} [\textit{ours}] &  &  \\ \cline{1-1}
FT+\textit{cDPO} [\textit{ours}] &  &  \\ \cline{1-1}
FT+\textit{SimPO} [\textit{ours}] &  &  \\ \hline
\end{tabular}
\end{table}

\subsubsection{Experimental Settings.}
To comprehensively assess the performance of our approach and the impact of the resulting dataset, we established a series of experimental evaluation settings according to Section~\ref{section:method-evaluation-experimental}, as shown in Table \ref{tab:experiment-settings}.

Model fine-tuning and alignment were conducted using dual NVIDIA A100 80GB GPUs with BF16 precision. For fine-tuning, we used a micro-batch size of 2 with 16 gradient accumulation steps (global batch size 32) over 10 epochs. The alignment process employed LoRA under GPU constraints, using a micro-batch size of 1 with 64 accumulation steps (global batch size 64) over 5 epochs. Shared hyperparameters included: learning rate of \verb|1e-5|, max sequence length of 4096 tokens, and LoRA configuration ($r=8$, $\alpha=16$, dropout=0.05).

\begin{figure}[thb]
    \centering
    \includegraphics[width=0.85\linewidth]{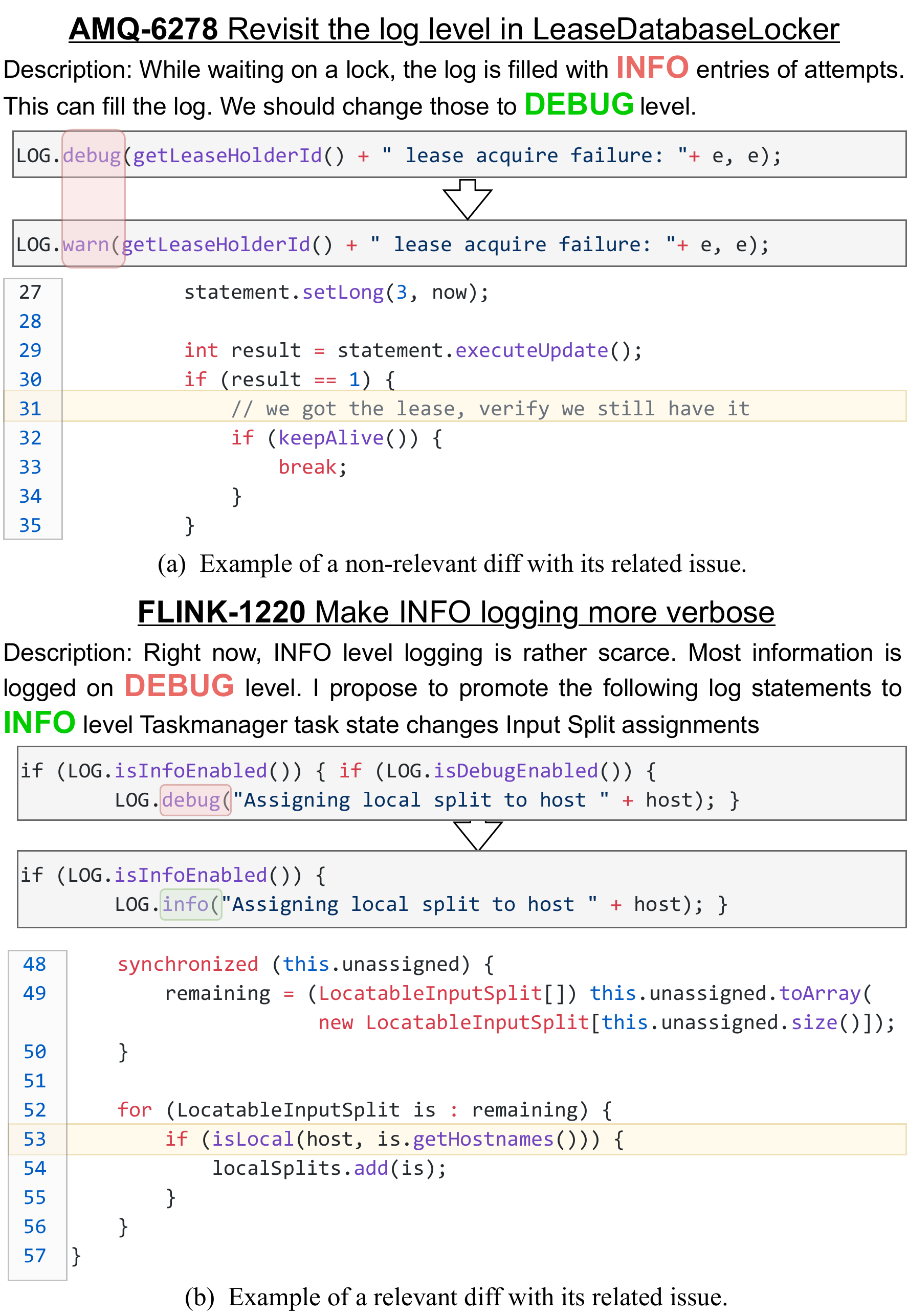}
    \caption{Example of non-relevant and relevant diffs with their related issues.}
    \label{fig:countersamples-combined}
\end{figure}

\section{Result Analysis}\label{section:eval}

In this section, we analyze the results obtained.

\subsection{Validation with Logging Practices (RQ1)}

Cohen's kappa coefficient calculated based on the overlapped 174 samples is 0.886, which shows that the two researchers are fairly consistent in evaluating the results. Yet after joint discussions, we finally only filtered 18 non-relevant diffs, which means that the overwhelming majority of our diffs are relevant to the related issues (808/826 = 97.8\%), and their quality can be guaranteed by the quality of well-discussed issues in the Apache community. Evidently, minor modifications like fixing typos and merely paraphrasing static text are considered trivial and lack instructional value for understanding effective logging practices.

Nevertheless, the filtered diffs are only not that relevant to the issue rather than necessarily totally irrelevant to the issue. For example, as Fig. \ref{fig:countersamples-combined} (a) shows, in the call of adjusting the redundant INFO level log statement to DEBUG, the diff reversely raises the level of a log statement from DEBUG to WARN. In the context of the code, this modification may possibly be reasonable as the importance of the \textit{acquire failure} message is worth the level of WARN, but anyway, it is different from the appeal of the issue. Consequently, the diff is excluded in our \AucadLog\ dataset.

In contrast, in the upcoming example shown in Fig. \ref{fig:countersamples-combined} (b), the diff is congruent with the claim of the issue that more information should be logged in INFO level rather than the DEBUG level, as a result, we consider it relevant to the issue it related to.

\begin{table*}[tb]
\caption{Experimental evaluation results of the LANCE testing set and LogBench-T dataset.}
\label{tab:experiment-result}
\footnotesize
\centering
\begin{tabular}{lllllllll}
\hline
\textbf{Method (Backbone)} & \textbf{PA} & \textbf{LA} &  \textbf{Adj. LA} & \textbf{MA}  & \textbf{BLEU-DM} & \textbf{VP}  & \textbf{VR} & \textbf{VF1} \\ \hline

\multicolumn{9}{l}{\textit{\textbf{Results of the LANCE testing set}:}} \\ \hline
LANCE~\cite{LANCE} & 0.546 & 0.378 &  0.546 & 0.100  & 0.110 & 0.271  & 0.253 & 0.254 \\ 
FastLog~\cite{xie2023generate} & 0.527 & 0.410 & 0.505 & 0.156  & 0.174 & 0.228  & 0.234 & 0.224 \\ 
UniLog~\cite{UniLog} (Llama 3.1) & \begin{tabular}[c]{@{}r@{}}0.523 \end{tabular} & \begin{tabular}[c]{@{}r@{}}0.522\end{tabular} & \begin{tabular}[c]{@{}r@{}}0.674 \end{tabular} & \begin{tabular}[c]{@{}r@{}}0.239 \end{tabular} & \begin{tabular}[c]{@{}r@{}}0.255 \end{tabular} & \begin{tabular}[c]{@{}r@{}}0.356 \end{tabular} & \begin{tabular}[c]{@{}r@{}}0.348 \end{tabular} & \begin{tabular}[c]{@{}r@{}}0.345 \end{tabular} \\
UniLog~\cite{UniLog} (Magicoder) & 0.116 &	0.213 &	0.315 &	0.077 &	0.081 &	0.121 &	0.119 &	0.119 \\ \hline
Llama 3.1 & 0.380 & 0.473 & 0.657 & 0.188 & 0.226 & 0.364 & 0.357 & 0.350  \\
FT (Llama 3.1) & \begin{tabular}[c]{@{}r@{}}0.571 (+ 50.5\%)\end{tabular} & \begin{tabular}[c]{@{}r@{}}0.570 (+20.5\%)\end{tabular} & \begin{tabular}[c]{@{}r@{}}0.718 (+\ 9.2\%)\end{tabular} & \begin{tabular}[c]{@{}r@{}}0.264 (+40.2\%)\end{tabular} & \begin{tabular}[c]{@{}r@{}}0.281 (+24.4\%)\end{tabular} & \begin{tabular}[c]{@{}r@{}}0.474 (+30.1\%)\end{tabular} & \begin{tabular}[c]{@{}r@{}}0.449 (+25.8\%)\end{tabular} & \begin{tabular}[c]{@{}r@{}}0.451 (+28.9\%)\end{tabular} \\
FT+\textit{DPO} (Llama 3.1) [\textit{ours}]& \begin{tabular}[c]{@{}r@{}}0.634 (+67.2\%)\end{tabular} & \begin{tabular}[c]{@{}r@{}}0.581 (+22.8\%)\end{tabular} & \begin{tabular}[c]{@{}r@{}}0.743 (+13.0\%)\end{tabular} & \begin{tabular}[c]{@{}r@{}}0.273 (+45.0\%)\end{tabular} & \begin{tabular}[c]{@{}r@{}}0.292 (+29.4\%)\end{tabular} & \begin{tabular}[c]{@{}r@{}}0.484 (+33.1\%)\end{tabular} & \begin{tabular}[c]{@{}r@{}}0.467 (+30.7\%)\end{tabular} & \begin{tabular}[c]{@{}r@{}}0.465 (+32.9\%)\end{tabular} \\
FT+\textit{cDPO} (Llama 3.1) [\textit{ours}] & \begin{tabular}[c]{@{}r@{}}\textbf{0.635} (+67.2\%)\end{tabular} & \begin{tabular}[c]{@{}r@{}}0.586 (+23.8\%)\end{tabular} & \begin{tabular}[c]{@{}r@{}}0.738 (+12.4\%)\end{tabular} & \begin{tabular}[c]{@{}r@{}}0.274 (+45.6\%)\end{tabular} & \begin{tabular}[c]{@{}r@{}}0.291 (+29.1\%)\end{tabular} & \begin{tabular}[c]{@{}r@{}}0.486 (+33.7\%)\end{tabular} & \begin{tabular}[c]{@{}r@{}}0.466 (+30.6\%)\end{tabular} & \begin{tabular}[c]{@{}r@{}}0.466 (+33.2\%)\end{tabular} \\
FT+\textit{SimPO} (Llama 3.1) [\textit{ours}] & \begin{tabular}[c]{@{}r@{}}\textbf{0.635} (+67.4\%)\end{tabular} & \begin{tabular}[c]{@{}r@{}}\textbf{0.591} (+25.0\%)\end{tabular} & \begin{tabular}[c]{@{}r@{}}\textbf{0.745} (+13.3\%)\end{tabular} & \begin{tabular}[c]{@{}r@{}}\textbf{0.275} (+46.1\%)\end{tabular} & \begin{tabular}[c]{@{}r@{}}\textbf{0.293} (+29.9\%)\end{tabular} & \begin{tabular}[c]{@{}r@{}}\textbf{0.488} (+34.2\%)\end{tabular} & \begin{tabular}[c]{@{}r@{}}\textbf{0.468} (+31.1\%)\end{tabular} & \begin{tabular}[c]{@{}r@{}}\textbf{0.468} (+33.8\%)\end{tabular} \\ 
\hline
Magicoder & 0.303 & 0.432 & 0.559 & 0.148 & 0.180 & 0.285 & 0.275 & 0.272  \\
FT (Magicoder) & \begin{tabular}[c]{@{}r@{}}0.493 (+62.4\%)\end{tabular} & \begin{tabular}[c]{@{}r@{}}0.429 (-\ 0.5\%)\end{tabular} & \begin{tabular}[c]{@{}r@{}}0.534 (-\ 4.4\%)\end{tabular} & \begin{tabular}[c]{@{}r@{}}0.182 (+23.3\%)\end{tabular} & \begin{tabular}[c]{@{}r@{}}0.202 (+11.8\%)\end{tabular} & \begin{tabular}[c]{@{}r@{}}0.343 (+20.0\%)\end{tabular} & \begin{tabular}[c]{@{}r@{}}0.326 (+18.3\%)\end{tabular} & \begin{tabular}[c]{@{}r@{}}0.325 (+19.6\%)\end{tabular} \\
FT+\textit{DPO} (Magicoder) [\textit{ours}]  & \begin{tabular}[c]{@{}r@{}}0.507 (+67.3\%)\end{tabular} & \begin{tabular}[c]{@{}r@{}}\textbf{0.444} (+\ 2.8\%)\end{tabular} & \begin{tabular}[c]{@{}r@{}}0.552 (-\ 1.2\%)\end{tabular} & \begin{tabular}[c]{@{}r@{}}0.189 (+27.7\%)\end{tabular} & \begin{tabular}[c]{@{}r@{}}0.210 (+16.5\%)\end{tabular} & \begin{tabular}[c]{@{}r@{}}0.351 (+23.0\%)\end{tabular} & \begin{tabular}[c]{@{}r@{}}0.334 (+21.5\%)\end{tabular} & \begin{tabular}[c]{@{}r@{}}0.334 (+22.7\%)\end{tabular} \\
FT+\textit{cDPO} (Magicoder) [\textit{ours}] & \begin{tabular}[c]{@{}r@{}}0.510 (+68.0\%)\end{tabular} & \begin{tabular}[c]{@{}r@{}}0.443 (+\ 2.6\%)\end{tabular} & \begin{tabular}[c]{@{}r@{}}0.554 (-\ 0.9\%)\end{tabular} & \begin{tabular}[c]{@{}r@{}}\textbf{0.190} (+28.8\%)\end{tabular} & \begin{tabular}[c]{@{}r@{}}\textbf{0.211} (+17.1\%)\end{tabular} & \begin{tabular}[c]{@{}r@{}}\textbf{0.355} (+24.5\%)\end{tabular} & \begin{tabular}[c]{@{}r@{}}\textbf{0.339} (+23.1\%)\end{tabular} & \begin{tabular}[c]{@{}r@{}}\textbf{0.338} (+24.3\%)\end{tabular} \\
FT+\textit{SimPO} (Magicoder) [\textit{ours}]  & \begin{tabular}[c]{@{}r@{}}\textbf{0.511} (+68.3\%)\end{tabular} & \begin{tabular}[c]{@{}r@{}}\textbf{0.444} (+\ 2.9\%)\end{tabular} & \begin{tabular}[c]{@{}r@{}}0.554 (-\ 0.9\%)\end{tabular} & \begin{tabular}[c]{@{}r@{}}0.188 (+27.6\%)\end{tabular} & \begin{tabular}[c]{@{}r@{}}\textbf{0.211} (+16.8\%)\end{tabular} & \begin{tabular}[c]{@{}r@{}}0.354 (+23.9\%)\end{tabular} & \begin{tabular}[c]{@{}r@{}}0.337 (+22.5\%)\end{tabular} & \begin{tabular}[c]{@{}r@{}}0.336 (+23.7\%)\end{tabular} \\ \hline

\multicolumn{9}{l}{\textit{\textbf{Results of the LogBench-T dataset}:}} \\ \hline
LANCE~\cite{LANCE} & 0.506 & 0.356 &  0.539 & 0.059  & 0.071 & 0.341  & 0.333 & 0.332 \\ 
FastLog~\cite{xie2023generate} & 0.493 & 0.429 &  0.571 & 0.095  & 0.118 & 0.330  & 0.339 & 0.329 \\ 
UniLog~\cite{UniLog} (Llama 3.1) & 0.347 &	0.442 &	0.671 &	0.106 &	0.132 &	0.334 &	0.333 &	0.330 \\ 
UniLog~\cite{UniLog} (Magicoder) & 0.158 &	0.210 &	0.345 &	0.041 &	0.049 &	0.141 &	0.140 &	0.139 \\ \hline
Llama 3.1 & 0.411 & 0.503 & 0.714 & 0.100 & 0.144 & 0.396 & 0.393 & 0.388 \\
FT (Llama 3.1) & \begin{tabular}[c]{@{}r@{}}0.592 (+44.1\%)\end{tabular} & \begin{tabular}[c]{@{}r@{}}0.529 (+\ 5.3\%)\end{tabular} & \begin{tabular}[c]{@{}r@{}}0.708 (-\ 0.9\%)\end{tabular} & \begin{tabular}[c]{@{}r@{}}0.129 (+28.9\%)\end{tabular} & \begin{tabular}[c]{@{}r@{}}0.154 (+\ 7.2\%)\end{tabular} & \begin{tabular}[c]{@{}r@{}}0.441 (+11.4\%)\end{tabular} & \begin{tabular}[c]{@{}r@{}}0.426 (+\ 8.4\%)\end{tabular} & \begin{tabular}[c]{@{}r@{}}0.427 (+10.0\%)\end{tabular} \\
FT+\textit{DPO} (Llama 3.1) [\textit{ours}] & \begin{tabular}[c]{@{}r@{}}0.639 (+55.4\%)\end{tabular} & \begin{tabular}[c]{@{}r@{}}0.528 (+\ 5.0\%)\end{tabular} & \begin{tabular}[c]{@{}r@{}}0.763 (+\ 6.9\%)\end{tabular} & \begin{tabular}[c]{@{}r@{}}0.138 (+37.6\%)\end{tabular} & \begin{tabular}[c]{@{}r@{}}0.167 (+16.4\%)\end{tabular} & \begin{tabular}[c]{@{}r@{}}0.450 (+13.7\%)\end{tabular} & \begin{tabular}[c]{@{}r@{}}0.438 (+11.3\%)\end{tabular} & \begin{tabular}[c]{@{}r@{}}0.437 (+12.6\%)\end{tabular} \\
FT+\textit{cDPO} (Llama 3.1) [\textit{ours}] & \begin{tabular}[c]{@{}r@{}}0.640 (+55.7\%)\end{tabular} & \begin{tabular}[c]{@{}r@{}}0.525 (+\ 4.3\%)\end{tabular} & \begin{tabular}[c]{@{}r@{}}0.767 (+\ 7.4\%)\end{tabular} & \begin{tabular}[c]{@{}r@{}}\textbf{0.140} (+39.9\%)\end{tabular} & \begin{tabular}[c]{@{}r@{}}\textbf{0.168} (+16.8\%)\end{tabular} & \begin{tabular}[c]{@{}r@{}}0.459 (+15.8\%)\end{tabular} & \begin{tabular}[c]{@{}r@{}}0.445 (+13.3\%)\end{tabular} & \begin{tabular}[c]{@{}r@{}}0.445 (+14.6\%)\end{tabular} \\
FT+\textit{SimPO} (Llama 3.1) [\textit{ours}] & \begin{tabular}[c]{@{}r@{}}\textbf{0.642} (+56.1\%)\end{tabular} & \begin{tabular}[c]{@{}r@{}}\textbf{0.536} (+\ 6.5\%)\end{tabular} & \begin{tabular}[c]{@{}r@{}}\textbf{0.769} (+\ 7.7\%)\end{tabular} & \begin{tabular}[c]{@{}r@{}}\textbf{0.140} (+39.8\%)\end{tabular} & \begin{tabular}[c]{@{}r@{}}\textbf{0.168} (+16.6\%)\end{tabular} & \begin{tabular}[c]{@{}r@{}}\textbf{0.464 } (+17.1\%)\end{tabular} & \begin{tabular}[c]{@{}r@{}}\textbf{0.451} (+14.7\%)\end{tabular} & \begin{tabular}[c]{@{}r@{}}\textbf{0.450} (+16.0\%)\end{tabular} \\ \hline
Magicoder & 0.250 & 0.409 & 0.492 & 0.072 & 0.107 & 0.282 & 0.276 & 0.274 \\
FT (Magicoder) & \begin{tabular}[c]{@{}r@{}}0.475 (+90.0\%)\end{tabular} & \begin{tabular}[c]{@{}r@{}}0.432 (+\ 5.8\%)\end{tabular} & \begin{tabular}[c]{@{}r@{}}0.573 (+16.5\%)\end{tabular} & \begin{tabular}[c]{@{}r@{}}0.098 (+35.4\%)\end{tabular} & \begin{tabular}[c]{@{}r@{}}0.124 (+16.2\%)\end{tabular} & \begin{tabular}[c]{@{}r@{}}0.338 (+20.1\%)\end{tabular} & \begin{tabular}[c]{@{}r@{}}0.328 (+19.1\%)\end{tabular} & \begin{tabular}[c]{@{}r@{}}0.327 (+19.5\%)\end{tabular} \\
FT+\textit{DPO} (Magicoder) [\textit{ours}]  & \begin{tabular}[c]{@{}r@{}}\textbf{0.495} (+97.8\%)\end{tabular} & \begin{tabular}[c]{@{}r@{}}0.448 (+\ 9.6\%)\end{tabular} & \begin{tabular}[c]{@{}r@{}}\textbf{0.595} (+20.8\%)\end{tabular} & \begin{tabular}[c]{@{}r@{}}\textbf{0.108} (+49.5\%)\end{tabular} & \begin{tabular}[c]{@{}r@{}}0.131 (+22.8\%)\end{tabular} & \begin{tabular}[c]{@{}r@{}}0.347 (+23.1\%)\end{tabular} & \begin{tabular}[c]{@{}r@{}}0.337 (+22.1\%)\end{tabular} & \begin{tabular}[c]{@{}r@{}}0.335 (+22.4\%)\end{tabular} \\
FT+\textit{cDPO} (Magicoder) [\textit{ours}]  & \begin{tabular}[c]{@{}r@{}}0.493 (+97.0\%)\end{tabular} & \begin{tabular}[c]{@{}r@{}}0.442 (+\ 8.3\%)\end{tabular} & \begin{tabular}[c]{@{}r@{}}0.593 (+20.5\%)\end{tabular} & \begin{tabular}[c]{@{}r@{}}0.106 (+47.0\%)\end{tabular} & \begin{tabular}[c]{@{}r@{}}\textbf{0.132} (+23.3\%)\end{tabular} & \begin{tabular}[c]{@{}r@{}}0.350 (+24.2\%)\end{tabular} & \begin{tabular}[c]{@{}r@{}}\textbf{0.340} (+23.2\%)\end{tabular} & \begin{tabular}[c]{@{}r@{}}\textbf{0.338} (+23.4\%)\end{tabular} \\
FT+\textit{SimPO} (Magicoder) [\textit{ours}]  & \begin{tabular}[c]{@{}r@{}}0.493 (+97.2\%)\end{tabular} & \begin{tabular}[c]{@{}r@{}}\textbf{0.449} (+\ 9.9\%)\end{tabular} & \begin{tabular}[c]{@{}r@{}}\textbf{0.595} (+20.8\%)\end{tabular} & \begin{tabular}[c]{@{}r@{}}0.105 (+45.2\%)\end{tabular} & \begin{tabular}[c]{@{}r@{}}0.130 (+21.5\%)\end{tabular} & \begin{tabular}[c]{@{}r@{}}\textbf{0.351} (+24.4\%)\end{tabular} & \begin{tabular}[c]{@{}r@{}}\textbf{0.340} (+23.3\%)\end{tabular} & \begin{tabular}[c]{@{}r@{}}\textbf{0.338} (+23.6\%)\end{tabular} \\ \hline
\end{tabular}
\vskip 0.3em
\raggedright
\textsuperscript{\dag} Llama 3.1 model adopts \(\beta=0.5\) for \textit{DPO}, \(\beta=0.1\) for \textit{cDPO}, \(\beta=2.5\) for \textit{SimPO}; Magicoder model adopts \(\beta=0.5\) for \textit{DPO}, \(\beta=1.0\) for both \textit{cDPO} and \textit{SimPO}.
\end{table*}

\subsection{Model Enhancement via Alignment (RQ2)}

To address \textbf{RQ2}, we investigate how the \AucadLog\ dataset can improve the capabilities of LLMs for the task of log statement recommendation after post-training.

The results with the optimal alignment hyperparameters determined after multiple attempts presented in Table~\ref{tab:experiment-result} illustrate that the \AucadLog\ dataset plays a crucial role in enhancing model performance through effective alignment strategies. Both the Llama 3.1 and Magicoder models exhibit significant improvements when aligned with the \textit{cDPO} or \textit{SimPO} algorithms, compared with only fine-tuned models, demonstrating that the quality and importance of training data are instrumental in optimizing model capabilities.

\uline{\textit{Position}.} In the Llama 3.1 model, the SimPO algorithm achieves a remarkable 67.4\% increase in \textit{Position Accuracy} (\textit{PA}) compared with the original Llama 3.1, which indicates a substantial enhancement in the model’s ability to accurately recommend log statement positions. This improvement is critical, as accurate positioning of log statements directly impacts the clarity and utility of logs for developers. In contrast, the fine-tuned model only achieves an increase of 50.5\% and also shows improvements, but falls short of the performance gains achieved by alignment.
Correspondingly, there is a similar improvement for Magicoder, although not as significant as Llama 3.1, which illustrates the benefit of using \AucadLog\ dataset alignment.
Even on unseen code in the LogBench-T dataset, alignment has a positive impact, with the performance of Llama 3.1 increasing by 12\% compared to fine-tuning alone.
 
\uline{\textit{Level}.} The effectiveness of the \AucadLog\ dataset is further evidenced in verbose-level recommendations.   The Llama 3.1 model demonstrates increases in both \textit{Log Level} (\textit{LA}) and \textit{Adjusted Log Level} (\textit{Adj. LA}), with improvements of 25\% and 13.3\%, respectively. In contrast, the fine-tuned model also achieves 10.3\% and 2.5\% increase but it is less than the performance gains brought by alignment. On both Magicoder and LogBench-T datasets, evaluations of alignment models using the \AucadLog\ dataset showed similar performance (except \textit{Adj. LA} on Magicoder). This highlights the ability of the model to provide more contextually appropriate verbosity in log statements, which can enhance readability and understanding for users.

\uline{\textit{Content}.}
For the textual content in the log statements, we adopt \textit{MA} and \textit{BLEU-DM} for the accuracy of message text in the log statements, together with \textit{Variable Precision} (\textit{VP}), \textit{Variable Recall} (\textit{VR}), and \textit{Variable F1} (\textit{VF1}) for the accuracy of variables in log message. 
The results demonstrate that alignment techniques, especially \textit{SimPO}, bring significant improvements across these metrics. For \textit{MA} and \textit{BLEU-DM}, \textit{SimPO} consistently enhances the accuracy of message text, with \textit{BLEU-DM} scores indicating closer alignment between generated and reference text. This boost in \textit{BLEU-DM} scores suggests that the aligned models produce log messages with content that is both semantically and structurally more accurate.
In terms of variable accuracy, the \textit{VP}, \textit{VR}, and \textit{VF1} metrics show notable gains with alignment algorithms, with \textit{SimPO} providing the most substantial improvements. For example, \textit{SimPO} raises \textit{VP} by up to 34.2\% in Llama 3.1 on the \LANCE\ testing set, indicating enhanced precision in selecting correct variables. Similarly, \textit{VR} and \textit{VF1} improvements reflect better recall and balanced precision-recall performance, pointing to greater model flexibility and consistency in handling variable content. Overall, these results highlight that \textit{SimPO} and related alignment algorithms improve both the textual accuracy and variable handling in log statements, thus enabling more effective and reliable log recommendations.

In summary, the findings strongly indicate that the alignment facilitated by our \AucadLog\ dataset is pivotal in improving the performance of LLMs.
Notably, our implementation of UniLog shows a certain degree of performance improvement compared to its base model, Llama 3.1 (8B), aligning with the improvement reported in its original paper. This consistency suggests that our implementation is likely correct. However, in terms of impact, prompt engineering methods represented by UniLog do not demonstrate significant advantages at the 8B parameter scale of LLMs. In fact, post-training methods, especially alignment algorithms such as \textit{SimPO}, \textit{DPO}, and \textit{cDPO}, show more substantial performance gains, particularly excelling in key metrics of accuracy and consistency in generated content. This indicates that for models with smaller parameter scales, post-training alignment optimization may be more effective than prompt engineering.

\section{Discussion}\label{section:discussion}

In this section, we discuss some implications of our study.

\subsection{High-quality Dataset and its Automatic Construction}

High-quality datasets are crucial in the post-training of LLMs, especially in the alignment. Our work involves extracting the code before and after the resolution of log-related issues in open-source software systems. The extracted code includes both log statements and surrounding contextual code, serving as training corpora for generating high-quality log statements. The results have been exceptionally promising. Since these log-related issues have already been resolved, they represent optimal examples of where and what to log, two of the most critical aspects regarding log statements~\cite{Gu2023TSE,chen2021survey}. Otherwise, the issues would not have been raised and addressed. Moreover,  the value of these datasets extends beyond positive examples. They also contain instances of poor or sub-optimal log statements -- commonly referred to as negative examples. Applying alignment techniques, like \textit{DPO}~\cite{DPO}, \textit{cDPO}~\cite{cDPO}, and \textit{SimPO}~\cite{SimPO}, 
our approach and dataset have yielded outstanding results, with enhancing open-source LLMs outperforming the current state-of-the-art models in generating log statements. 
Another noteworthy point is that this method of dataset construction has to be automated given the huge labor costs. In this sense, our approach shows full potential, as the size of our dataset can be increased by incorporating more open source projects and log-related issues. To better implement the value of our work, we plan to make this dataset publicly available (see Section \textbf{Data Availability}) to allow interested researchers to replicate our study or explore further.

\subsection{The Significance of Automated Log Statement Generation}

Since large language models (LLMs) have been integrated into code completion and generation tools, their performance has gained substantial recognition, as demonstrated by the widespread adoption of tools like Copilot. These tools have attracted a considerable user base, and multiple reports suggest that they significantly enhance production efficiency~\cite{wermelinger2023using, dakhel2023github}. We contend that automated log statement generation is particularly critical within the realm of LLM-enabled code activities. Log statements are essential components of modern software systems, and if their creation does not keep pace with other business code, they could become a bottleneck that impedes overall software development productivity. 

Moreover, the quality of the data significantly impacts the fine-tuning of models.
Our constructed dataset effectively supports the fine-tuning of LLMs, thereby enhancing the generation of log statements. By fine-tuning open-source LLMs and subsequently utilizing the \AucadLog\ dataset to align models such as Llama 3.1 and Magicoder, we have demonstrated significant improvements in log statement generation. These fine-tuned models notably outperform UniLog and state-of-the-art fine-tuning approaches across multiple evaluation metrics.

Furthermore, our observations also indicate that our dataset can counteract the performance decline commonly associated with fine-tuning. Additionally, the degree of improvement achieved through alignment with our dataset surpasses that obtained with the fine-tuned models, underscoring the efficacy of high-quality data.

\section{Threats to Validity}\label{section:validity}

In this section, we discuss some factors that may potentially bring risks to the findings of the study.

\uline{Limited LLMs.}
Due to our constraint on GPU capabilities, we conducted fine-tuning experiments on Llama~3.1 and Magicoder,
which inevitably introduced some limitations in result generalization. However, \AUCAD\ is designed as a model-agnostic method to work with multiple LLMs, and we have also made our dataset and original response from LLMs publicly available (see Section \textbf{Data Availability}).
We encourage interested researchers to replicate our work on a broader range of LLMs and thoroughly validate the effectiveness of the proposed method.

\uline{Only involving log-related issues.}
Limiting our dataset to log-related issues imposes constraints on its diversity and the coverage of real-world logging scenarios. In practice, certain log statements and their associated code snippets are inherently sensible and may not be flagged as issues. Unfortunately, these valuable data points are excluded from our dataset. 
On the one hand, distilling high-quality datasets from issues is an effective way to fine-tune LLMs optimized for the log statement generation task. On the other hand, there is currently no automated method for extracting high-quality log statements and context code directly from open-source projects. Manual distillation, while theoretically possible, is too costly and time-consuming. Nevertheless, it is worth noting that the \AucadLog\ dataset has already covered all log-related issues in Apache (as discussed in Section~\ref{section:method-dataset}). As more data added to the dataset in the future, this problem should gradually be alleviated.

\uline{Replication bias.} Since the dataset, source code as well as the base LLM model are not accessible, we have to replicate the UniLog method, which may lead to replication bias. However,  the experimental results from our replicated UniLog show similar results to the metrics of the original paper~\cite{UniLog} in terms of the increasing trends of all the evaluation metrics, which to a large degree lowered the risk pertinent to replication bias.

\uline{About the level evaluation.}
To more accurately assess the log levels recommended by our model, we have introduced the metric \textit{Adjust Level Accuracy} (\textit{Adj. LA}), which focuses only on the log levels that need adjustment. However, it should be noted that this is not a perfect solution but rather a more reasonable compromise towards a pragmatic logging practice. 
We believe that further exploration is needed to develop more reasonable evaluation mechanisms.

\uline{About the content evaluation.} 
Borrowing from previous studies~\cite{LANCE, UniLog}, we also employed text similarity metrics, such as \textit{BLEU}~\cite{BLEU} and its variants (e.g., \textit{BLEU-DM}~\cite{shi2022evaluation}), to measure the quality of the generated log statements. However, the inherent content generation nature of LLMs poses challenges to objective and accurate evaluation using metrics like \textit{BLEU} and its variants, which also has been acknowledged in prior research~\cite{evtikhiev2023out}. 
To mitigate this threat, we extracted variables when evaluating log statement content. We employed metrics of variables for evaluations, e.g., \textit{Variable Precision} (\textit{VP}), \textit{Variable Recall} (\textit{VR}), and \textit{Variable F1 score} (\textit{VF1}), thereby providing multiple perspectives for evaluating the content of the log statements generated by LLM.
We acknowledge the validity risk associated with using this metric and will keep an open mind to tracking research advances to find better ways to evaluate the quality of log statement content.

\section{Conclusion}\label{section:conclusion}
Log data plays a pivotal role in the operation and maintenance of modern software systems. Logs are generated through log statements strategically placed throughout the software codebase. With the rapid advancements in Large Language Models (LLMs), there is potential for significantly improving code completion tasks, which includes the automated generation of log statements.
As a result, the automated generation of log statements has gained prominence in contemporary software development. Despite the progress in LLMs, their current capabilities in generating log statements still require enhancement to truly benefit software development processes~\cite{Li2024LogBench}.
Post-training with high-quality dataset is a widely accepted solution, however the huge cost of manual annotation makes this solution not very feasible in practice.
We address this gap by proposing \AUCAD, an automated method for constructing a high-quality post-training dataset, and by contributing \AucadLog, a custom-curated dataset comprising pre- and post-changed log statements and their corresponding code segments from real-world log-related issues in the open-source community.
By fine-tuning open-source LLMs then using \AucadLog\ to align them, Llama~3.1 and Magicoder, to devise exemplar models with the \AucadLog\ dataset, the fine-tuned models achieve significant improvements in log statement generation, notably outperforming UniLog and the state-of-the-art finetuning methods
in terms of multiple evaluation metrics.

Future research should consider extending the \AUCAD\ method to construct preference alignment datasets across a wider array of scenarios, incorporating various LLMs, diverse parameter scales, and multiple programming languages.   Additionally, exploring more nuanced code context ranges may offer further improvements in model performance.

\section*{Data Availability} \label{sec:DataAvailability}

Our \AucadLog\ dataset and the original response of LLMs are publicly available at \artifact.

\bibliographystyle{ACM-Reference-Format}
\bibliography{main}


\end{document}